# Built-in electric field induces polarization rotation in bilayer BiFeO$_3$/(Ba,Sr)TiO$_3$ thin films

A. G. Razumnaya[1], A. S. Mikheykin[1], D. V. Stryukov[2], A. S. Anokhin[2], A. V. Pavlenko[2], V. B. Shirokov[1,2], M. A. Kaliteevski[3], D. Mezzane[4], I. Lukyanchuk[5]

[1]*Faculty of Physics, Southern Federal University, Rostov-on-Don, 344090, Russia*
[2]*Southern Scientific Center of the Russian Academy of Sciences, pr. Chekhova 41, Rostov-on-Don, 344006 Russia*
[3]*ITMO University, Kronverkskiy pr. 49, St. Petersburg, 197101, Russian Federation*
[4]*LMCN, Cadi Ayyad University, Marrakesh, 40000, Morocco*
[5]*LPMC, University of Picardy Jules Verne, Amiens, 80080, France*

E-mail: agrazumnaya@sfedu.ru

**Abstract**

The crystal structure of BiFeO$_3$/Ba$_x$Sr$_{1-x}$TiO$_3$ (BFO/BST) heterostructures with $x = 0.2$, 0.6 and 0.8, grown on single-crystal MgO (001) substrate was investigated by x-ray diffraction and Raman spectroscopy in order to determine the influence of mismatch-induced strains and spontaneous polarization in BST buffer layers on BFO layers. The lattice parameter of the BFO layers was shown to decrease with increasing concentration of Ba ions, despite the increasing in-plain lattice parameters of tetragonal unit cells of BST layers. The rhombohedral angle of the crystal structure of BFO layers demonstrates an increase towards the ideal cubic perovskite structure with the appearance of the built-in electric field, induced by the spontaneous polarization in buffer layers. This result provides a remarkable tool for the control of polarization in BFO layers and other ferroelectric films in general, by changing the built-in electric field from ferroelectric buffer layer without changing a single crystal substrate.

## 1. Introduction

Strain engineering is a powerful tool to control the magnetic and ferroelectric properties of functional materials for application in nanoelectronic devices [6-8]. Within this approach, the properties of thin-film heterostructures are usually tuned by the substrate-imposed two-dimensional stress which has either tensile or compressive type. It is of particular interest to explore the strain-induced modifications for the most promising multiferroic compound, bismuth ferrite, BiFeO$_3$, (BFO) possessing both the antiferromagnetic and ferroelectric ordering at room temperature [1, 2] In the ferroelectric phase, occurring below the critical temperature $T_C = 1083$ K, the bulk BFO has the rhombohedral structure with the *R3c* symmetry [3]. Besides, below the Neel temperature, $T_N = 643$ K, the antiferromagnetic ordering occurs [4]. The possibility of the increasing of the magnetization and spontaneous polarization in the strained BFO films deposited on the SrTiO$_3$ (STO) substrate was clearly demonstrated in [5].

The inconvenience, frequently arising in strain engineering is the formation of dislocations inside the films that tend to compensate for the dynamically arising mismatch strain during the film deposition process. The latter is caused by the different thermal expansion of the film and substrate materials. Such defects modify the properties of the system in an uncontrolled way. This shortcoming can be overcome by exploring the damping effect of the cleverly selected buffer layer. Moreover, the buffer-layer approach can substantially improve the quality of the epitaxial deposition of the perovskite-on-perovskite heterostructures, providing the more precise tuning of the two-dimensional stresses in a thin film by changing the thickness and/or composition of the buffer layer. Application of the ferroelectric buffer layer opens additional possibilities for the manipulation of the system by the arising build-in electric field.

In Ref. [5] the intermediate buffer layer of SrRuO$_3$ (SRO) deposited in between the BFO film and STO substrate was used for these purposes. However, it appears even more advantageous to use the solid solution Ba$_x$Sr$_{1-x}$TiO$_3$ (BST$x$) since the lattice parameters of BFO and BST$x$ can be adjusted by variation of $x$. It was shown in Ref. [9] that the layer of BST0.25 of the thickness of 30 nm, grown by chemical deposition technique at the Pt/Si substrate does not affect the crystal structure of the deposited BFO film. At the same time, this layer reduces the leakage currents by three orders of magnitude with a slight decrease in magnetization. It was found that the formation of a multilayer BFO/BTO heterostructure on the (001)STO substrate provides the increase of magnetization [10]. In addition, the BFO films grown on STO substrates demonstrate the prominent current-voltage switching hysteresis, useful for applications [12]. Even better results for magnetization enhancement were achieved for the BFO/BST0.8 structure grown on (001)MgO substrate [11].

In this work we systematically study how the BST$x$ buffer layers affect the properties of the BFO films, using $x$ as a parameter that tunes the quality of the system. More specifically, we grow the BFO films on the MgO substrate, with intermediate BST$x$ ferroelectric buffer layers with $x = 0.2, 0.6$ and $0.8$ and of the thickness 100 nm. For detailed characterization of the BFO/BST$x$/MgO system, we use the Raman spectroscopy technique which is an important tool to test the crystal structure, lattice dynamics and spin-phonon interaction in multiferroics.

Our study is based on the comparative analysis of the previous Raman experiments for the BFO single-crystal and thin-film materials with our measurements for the BFO/BST$x$/MgO system. In Ref. [13] it was shown that all the first-order phonons appear in the Raman spectra of the BFO single crystal was below 600 cm$^{-1}$. Above 600 cm$^{-1}$, there is no significant features in the spectra besides a broad peak around 1250 cm$^{-1}$, which was interpreted as a higher-order phonon band. The disappearance of this band above 600-700 K was associated with the occurrence of the magnetic ordering [13]. Phonon anomalies in the epitaxial BFO thin film grown on the SRO substrate were studied in the temperature range from -192 up to 1000°C in Ref. [14]. Raman studies of the BFO films grown on (111)STO substrate in the temperature range of 300-800 K revealed the anomalous frequency shift of low-frequency optical modes of $A_1$-symmetry. This shift occurs due to the spin-phonon interaction and structural instability related to the modulation of the bond angle in the Fe-O-Fe chain near $T_N$ [15]. It was also discovered that Raman response in the frequency range of 1000-1300 cm$^{-1}$ is very sensitive to the antiferromagnetic phase transition, which indicates a strong spin-two-phonon interaction in BFO [16, 17].

Basing on our x-ray diffraction and Raman spectroscopy studies of BFO films grown on the MgO substrate with the intermediate ferroelectric buffer layers BST$x$ we demonstrate for the first time the possibility to control the direction of the polarization vector in the BFO thin films using the built-in electrostatic field arising due to the presence of polarization in the buffer BST layers.

## 2. Experimental details

The BiFeO$_3$/Ba$_x$Sr$_{1-x}$TiO$_3$ (BFO/BST$x$) heterostructures with $x = 0.2, 0.6$ and $0.8$ were grown on the (001)-oriented cubic single-crystal MgO substrate using rf sputtering using alternating focusing of the laser beam on stoichiometric BFO and BST$x$ targets. The total thickness of the grown bi-layer heterostructures was 200 nm, whereas the thickness of the BST$x$ buffer layers was 100 nm.

The structural perfection of the heterostructures, the unit cell parameters of the BFO and BST$x$ layers in a plane parallel to the substrate and perpendicular to the substrate, the average coherent scattering region size, and the average microdeformation in the heterostructures were determined using X-ray diffraction on a Rigaku Ultima IV diffractometer (CuK$_{\alpha 1}$ radiation, $\theta$-$2\theta$ and $\phi$ scan modes).

The surface microstructure of the BFO/BST$x$ heterostructures was studied using an Integra atomic force microscope (NT-MDT). The measurements were conducted in a semicontact mode using an NSG11 standard silicon cantilever. The analysis of the morphology shows that the average roughness of the film is only 0.13 nm, and the maximum height of the surface relief is less than 2 nm.

The Raman spectra were excited by polarized radiation from an argon laser with $\lambda = 514.5$ nm and were measured using a micro-Renishaw inVia Reflex spectrometer equipped with an Edge-filter for analysis of the spectra in the range from 50 cm$^{-1}$ and higher. The exciting radiation was focused on the sample using a Leica optical microscope, the focused beam diameter on the sample was 1-2 μm.

Raman spectra of the BFO/BST$x$ heterostructures were measured in the backscattering geometry on the samples which were exactly oriented in correspondence with the crystallographic axes of the MgO substrate: $X \parallel [100]$, $Y \parallel [010]$, and $Z \parallel [001]$. Polarized Raman spectra were recorded in the side-view backscattering geometry in which the wave vector of the incident beam was parallel to the substrate. The polarization of the incident and scattered light beams was parallel or perpendicular to the $Z$-axis of the film. The measured Raman spectra were corrected using the Bose-Einstein temperature factor $(\omega, T) = (e^{\frac{\hbar\omega}{kT}} - 1)^{-1}$.

## 3. Results and discussion

Fig. 1 presents the room-temperature x-ray diffraction $\theta$-$2\theta$ patterns of the studied BFO/BST$x$ heterostructures. The x-ray diffraction spectra contain only (*00l*) reflections from the BFO and BST$x$ layers and MgO substrate. No impurity phases were detected in the samples. The analysis of the x-ray diffraction patterns of the BFO/BST$x$ heterostructures shows that all the samples have a tetragonal symmetry within the experimental error.

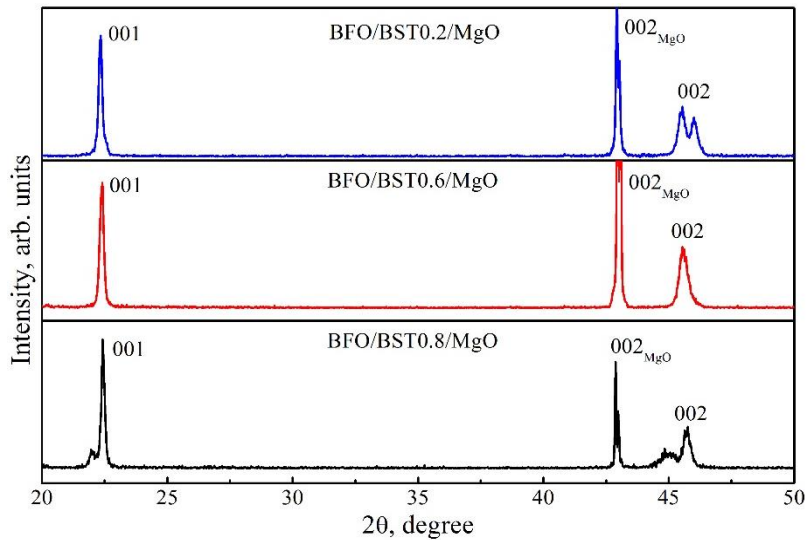

Fig. 1. Fragments of the x-ray $\theta$-$2\theta$ patterns of the BFO/BST$x$ heterostructures at room temperature.

The orientation relationships between the BFO and BST$x$ layers and MgO substrate have been determined by the x-ray diffraction in $\phi$ scan mode (Fig. 2) when the angle $\varphi$ varies from 0 to 360º. The $\varphi$-scans performed around the (113) reflections of the BFO/BST$x$ heterostructures and MgO substrates show that the diffraction patterns have four reflections spaced at 90º, confirming the parallel orientation of the bi-layer structures and substrates. The [100] and [010] axes correspond to the $\varphi$ angles equal to 0 and 90º, respectively. Therefore, there is only one azimuthal orientation of the film relative to the substrate: [001]$_{BFO/BST}$ ∥ [001]$_{MgO}$ and [100]$_{BFO/BST}$ ∥ [100]$_{MgO}$, [010]$_{BFO/BST}$ ∥ [010]$_{MgO}$ (Fig. 2). The rocking curve widths and $\varphi$-scans reveal that the vertical and azimuth misorientations of the BFO layers do not depend on the Ba concentration in buffer BST$x$ layers and were about ~1° and ~6.5°, respectively.

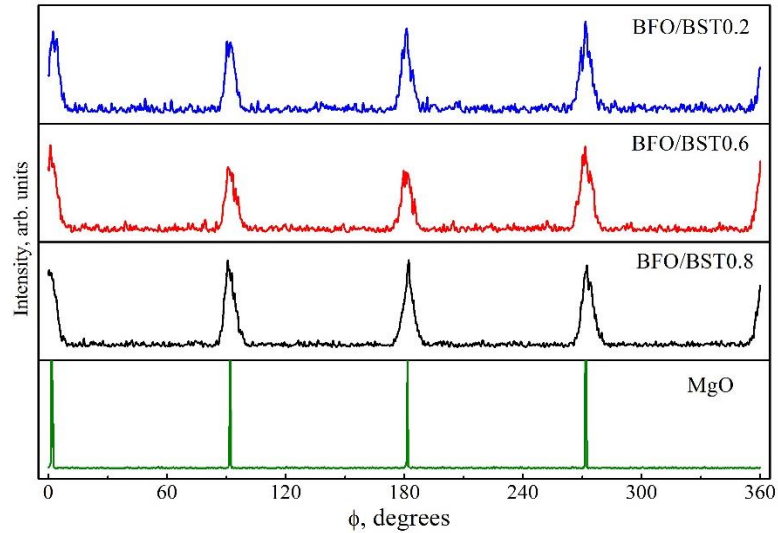

Fig. 2. $\varphi$-scans of the (113) reflections of the BFO/BST$x$ heterostructures and MgO substrates in the range of $\varphi$ angles from 0 to 360º.

The measured (*00l*) reflection profiles of the BFO/BST$x$ heterostructures are shown in Fig. 3. The out-of-plane lattice $c$-parameters of the BFO and BST layers were obtained by Gaussian function approximation of the (*00l*) reflections. The in-plane lattice $a$-parameters of the BFO/BST$x$ heterostructures were determined using asymmetric $\theta$-$2\theta$-scanning of the (113) reflections with rotation angle $\varphi$ (Fig. 4).

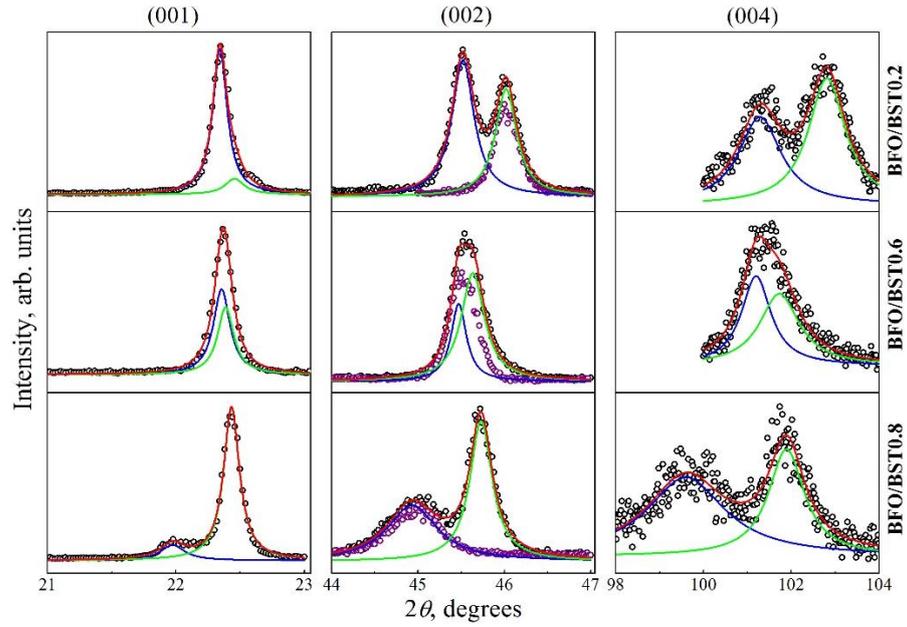

Fig. 3. Zoom of the θ-2θ diffraction patterns for the (001), (002) and (004) reflections of the studied BFO/BSTx heterostructures ($x$ = 0.2, 0.6 и 0.8) at room temperature. Experimental data are shown by open symbols, whereas solid lines correspond to the fitting profiles. The purple symbols show the (002) reflections of the single-component BST0.2, BST0.6 and BST0.8 films before deposition of the BFO layers.

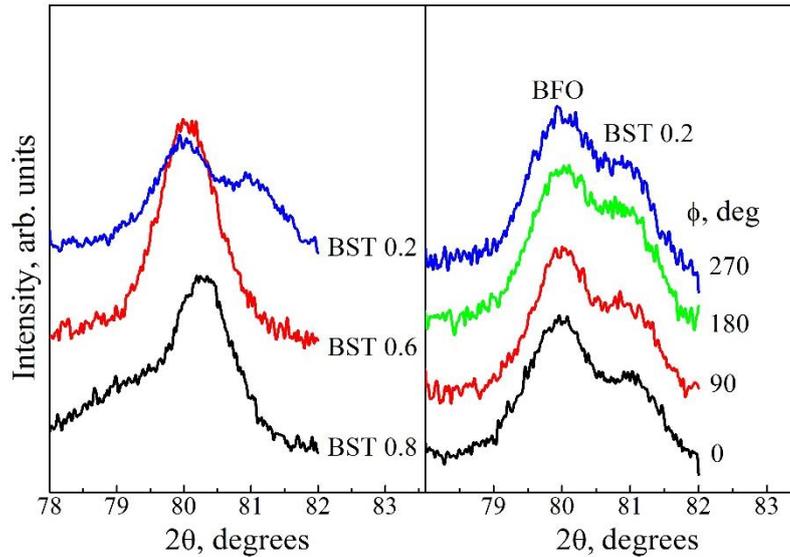

Fig. 4. Asymmetric θ-2θ-scanning of the (113) reflections of the BFO/BSTx heterostructures with rotation angle φ.

The results of the processing of x-ray diffraction patterns of the BFO/BSTx structures were collected in Table 1. Summarized data show that the $a$ and $c$ lattice parameters and cell volumes $V$ monotonically increase when the Ba concentration increase gradually in the buffer BSTx layers. The observed dependence $V(x)$ can be explained by the difference in the ionic radii $R_{Ba^{2+}}$ (Å) and $R_{Sr^{2+}}$ (Å) during the formation of the $Ba_xSr_{1-x}TiO_3$ solid solutions. The detailed analysis of the x-ray diffraction data shows that the BFO and BSTx layers demonstrate the tetragonal distortion $c/a > 1$. It worth noting that the degree of distortion of the BSTx layers increases with increasing of Ba concentration, while the structural distortion of the BFO layers remains virtually unchanged.

Note, that the bulk ceramic BSTx materials with $x$ = 0.2 and 0.6 compositions have the cubic $Pm3m$ space group with $a$ = 3.922 and 3.970 respectively, whereas the $Ba_{0.8}Sr_{0.2}TiO_3$ composition has the tetragonal *P4mm* space group with $a = 3.977$ Å and $c = 3.988$ Å [18, 19]. At the same time, the ceramic BFO sample has a rhombohedral structure with parameters $a = 3.965$ Å and $\alpha = 89.45°$. As concerns the thin-film structures, various structural distortions of BFO, depending on the growth conditions, composition and orientation of the substrate, composition and thickness of the buffer layers, can be obtained. The monoclinic [5, 14, 20, 21], orthorhombic [22, 23], tetragonal [24, 25], rhombohedral [23, 24, 26] and triclinic [27] structures were observed.

Table 1. Structural parameters of the BFO and BST$x$ layers, forming bi-layer BFO/BST$x$ heterostructures grown on a cubic (001) MgO substrate, and BST$x$ films before deposition of the BFO layer at room temperature.

| Heterostructures | BST layer | | $V_{BST}$, Å$^3$ | BFO layer | | | | $V_{BFO}$, Å$^3$ | BST$x$ |
|---|---|---|---|---|---|---|---|---|---|
| | $c$, Å | $a$, Å | | $a$, Å | $\alpha$, deg. | $a_{PC}$, Å | $\alpha_{PC}$, deg. | | $c/a$ |
| BFO/BST$_{0.2}$ | 3.940(12) | 3.916 (75) | 60.42 (2.50) | 5.726(3) | 58.14(5) | 4.049(2) | 88.39(4) | 127.12(22) | 1.006(22) |
| BFO/BST$_{0.6}$ | 3.993(4) | 3.935 (60) | 61.83(1.95) | 5.683(6) | 58.66(12) | 4.018(4) | 88.84(11) | 125.12(49) | 1.015(17) |
| BFO/BST$_{0.8}$ | 4.034(8) | 3.947 (97) | 62.84 (3.21) | 5.670(1) | 58.82(3) | 4.009(1) | 88.98(3) | 125.45(11) | 1.022(27) |

For investigated here BFO/BST$x$ system the diffraction study show (Figs. 1 and 2) that the coupling between the layers in heterostructures is sufficiently strong so their orientations with respect to the crystallographic directions of the substrate completely coincide. The crystal structure of the BST layer behaves predictably with increasing the Ba concentration: the *a* parameter of the tetragonal unit cell increases (see insert in Fig.5), the parameter *c* and the ratio *c*/*a* also increase due to the occurrence of spontaneous polarization (Tables 1 and 2). Since the lattice parameters decrease with increasing of the Sr concentration in the BST$x$ layers, one expects that the buffer BST$x$ layers with $x < 0.5$, in particular, the BFO/BST0.2 will create the two-dimensional compressive stresses in the BFO layer, whereas the BST$x$ layers with $x > 0.5$, in particular, the BST0.6 and BST0.8 will produce the tensile stresses in the BFO layers. In both cases, the rhombohedral structure of the crystal lattice of the BFO layers is expected as in a bulk sample (left part of Fig. 6).

According to the diffraction data, the parameter *a* of the rhombohedral unit cell of BFO decreases and the parameter *a* of BST buffer layer increases when *x* increases. Such dependence of lattice parameters (Fig. 5) explains the increase of the rhombohedral angle α in the BFO layer on the Ba concentration *x*. The crystal structure of BFO layers becomes less distorted, approaching to the cubic lattice with $\alpha = 60°$. The changes of the rhombohedral angle led to the polarization vector rotation in the BFO layers. This can be explained by the influence of the electric field produced by the electric dipoles in the BST layers. This field effects on the polarization the BFO layers as shown in the right part of Fig. 6.

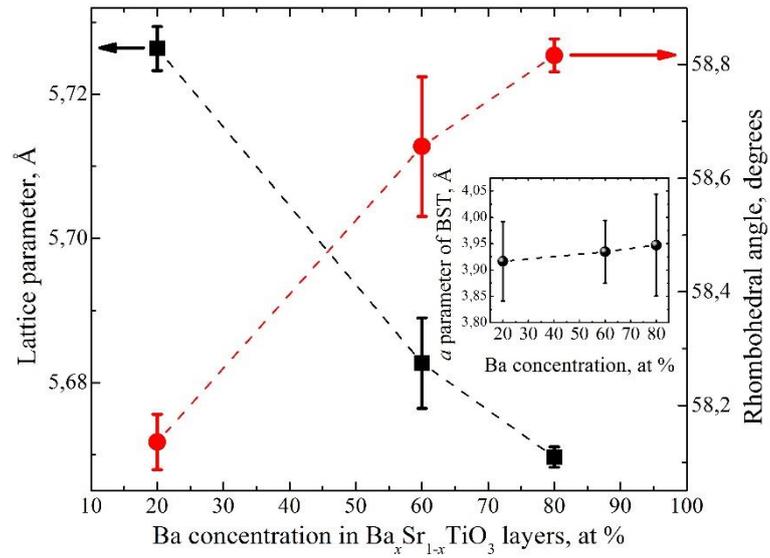

Fig 5. Lattice parameter (squares) and rhombohedral angle (flat circles) of the BFO layers crystal structure as a function of the Ba concentration in the BST layers. The inserted panel shows the in-plane *a* lattice parameter as a function of the Ba concentration in the BST layers.

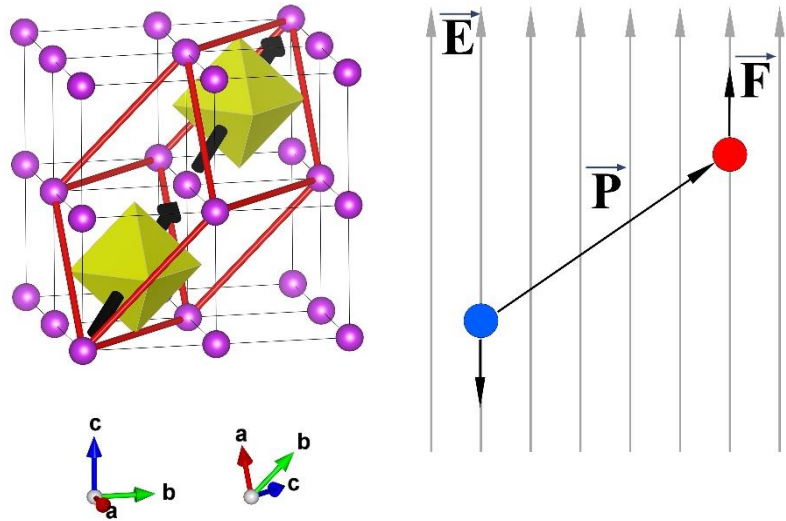

Fig 6. The rhombohedral unit cell of the crystal structure of BFO layers (left) and rotation of polarization vector of BFO layers under the electrostatic build-in field from poled BST layers (right).

The investigation of the surface morphology of BFO/BST*x* heterostructures by atomic force microscopy (Fig. 7) demonstrated that the BFO layers were grown by the 3D island growth mechanism rather than by the layer-by-layer mechanism, the islands being strictly oriented with respect to each other and substrate as follows from $\varphi$-scans of (Fig. 2).

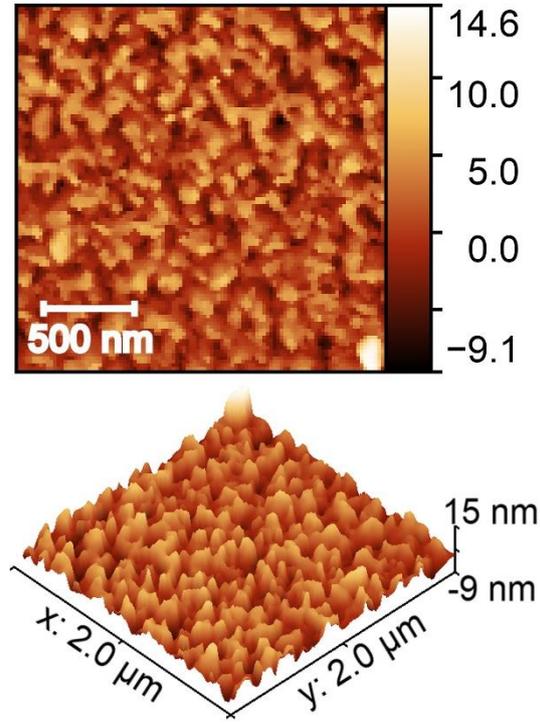

Fig. 7. Microrelief of a surface fragment of the BFO, grown on a BST0.8 buffer layer by 3D island growth mechanism: (a) two-dimensional image and (b) three-dimensional image.

Using the buffer layers and substrates with various Ba composition allows manipulation the strain in the film, arising from the lattice mismatch of the film-substrate lattice constants and, hence vary the ferroelectric properties of these structures. Comparing the obtained lattice parameters for studied heterostructures (Table 1) with the lattice parameters of the bulk materials of the same composition, one can obtain the microstrains of the lattice, arising due to the lattice mismatch between film and substrate (Table 2). The corresponding misfit strains in the plane of the substrate, $\varepsilon_{11}$, and along the substrate normal, $\varepsilon_{33}$, are defined as:

$$\varepsilon_{11} = (a_f - a_b)/a_b,$$
$$\varepsilon_{33} = (c_f - c_b)/c_b,$$

where $c_f$ and $a_f$ are the unit cell parameters of the film, $c_b$ and $a_b$ are the unit cell parameters of the bulk material [28]. It was also found that the unit cells of the layers compress in the plane of conjugation and stretch in the perpendicular direction. Moreover, we observed the decreasing of the unit cell volumes in comparison with the bulk material ($V/V_b < 1$). The value of strain decreases in the BFO layers with simultaneous relaxation of the unit cell volume $V$ with decreasing of the Ba concentration in the buffer BST$x$ layers.

Using the dependence of the degree of tetragonality $c/a$-1 as a function of the misfit strain $u_m$, [29] we find the misfit strain and polarization of BST$x$ layers at different $x$, see Table 2. It is confirmed that the polarization in the BST0.2 layer is absent and the strain in the plane coincides with $\varepsilon_{11}$. It worth noting that the parameters given in Table 2 were calculated using a model of a mechanically strained film. For the real heterostructure, the BST0.2 layer is clamped by the MgO substrate and BFO film. Note, that due to large differences in the bandgap width, an electric double layer is formed at the interface between MgO substrate and BST layers [30]. The parameters of such in-built electric layer depend on the deformation of the film that can be controlled by the technology of deposition.

Table 2. The values of the degree of tetragonality ($c/a$-1), the misfit strain of the films with respect to the bulk sample $u_m$, and the spontaneous polarization $p$ for the BST$x$ layers in the BFO/BST$x$/(001)MgO heterostructures.

| Heterostructures | BST layer | | |
|---|---|---|---|
| | $c/a$-1 | $u_m$ | $p$, C/m$^2$ |
| BFO/BST$_{0.2}$/MgO | 0.006 | -0.004 | 0.0 |
| BFO/BST$_{0.6}$/MgO | 0.015 | -0.007 | 0.26 |
| BFO/BST$_{0.8}$/MgO | 0.022 | -0.010 | 0.30 |

The room-temperature polarized Raman spectra of the BFO/BST$x$ samples in the frequency range of 50-1500 cm$^{-1}$ at different scattering geometries are shown in Fig. 8. The contribution from the top BFO layer dominants in the all spectra when the thicknesses of the layers are ~100 nm, that is consistent with a study reported in [20]. As mentioned above, the BFO layer exhibits numerous structural modifications. Rhombohedral, $R3c$, tetragonal, $P4mm$, and monoclinic, $B_b$, distortions of the BFO films lead to 13, 8 and 27 Raman active modes, decomposed as:

$$\Gamma_{Rhombohedral}\ (C_{3v}) = 4A_1 + 9E,$$
$$\Gamma_{Tetragonal}\ (C_{4v}) = 3A_1 + B_1 + 4E,$$
$$\Gamma_{Monoclinic}\ (C_s) = 13A' + 14A''.$$

The selection rules for the given BFO structures in different scattering geometries are presented in Table 3. The polarized Raman spectra of the BFO/BST$x$ heterostructures for different scattering geometries are presented in Fig. 8. For comparison, we also show the Raman spectra of the BFO thin films. Above 600 cm$^{-1}$ the all spectra contain bands corresponding to the first-order phonon modes of the bismuth ferrite. A detailed assignment of phonon modes was performed earlier in [14]. Note, that the spectra of studied BFO/BST$x$ heterostructures in $XX$ and $YY$ scattering geometries are almost identical and contain vibrational modes with the same frequencies, the most intense lines are 75, 147, 170, 228, 475, 525, 630, 1220 and 1280 cm$^{-1}$.

Table 3. Selection rules of Raman active modes for a rhombohedral R, a tetragonal T, and monoclinic M crystal structures in different scattering geometries. Line N(Raman) presents the total number of Raman active modes corresponding to the depolarized spectrum.

| Scattering Geometry | Space group | | |
|---|---|---|---|
| | R ($R3c$) | T ($P4mm$) | M ($B_b$) |
| N(Raman) | $4A_1 + 9E$ | $3A_1 + B_1 + 4E$ | $13A' + 14A''$ |
| $Z(YY)\bar{Z}$ | $A_1$ and E | $A_1$ and $B_1$ | $A'$ |
| $Z(YX)\bar{Z}$ | E | – | $A''$ |
| $Y(XX)\bar{Y}$ | $A_1$ и E | $A_1$ и $B_1$ | $A'$ |
| $Y(XZ)\bar{Y}$ | E | E | $A'$ |

The frequencies of the Raman lines in the spectra of the BFO/BST$x$ heterostructure slightly different from those, observed in [14] due to the structural distortions of the BFO and BST layers. In the case of the tetragonal symmetry, $P4mm$, when the polar $Z$-axis is normal to the substrate, the $YX$ spectrum is forbidden by selection rules [31]. Nevertheless, we observe the low-intensity Raman response in the $YX$ spectra of the studied BFO/BST$x$ heterostructures (Fig. 8), which eliminates the tetragonal symmetry of the layers. Moreover, the frequencies of the lines in the $YX$ spectra coincide with the corresponding lines in the $XZ$ spectra. The similar lines in the spectra of the $YX$ and $XZ$ scattering geometries are inconsistent with the selection rules for the monoclinic symmetry of the crystal structure of the BFO layers. As mentioned above, the BFO layers were grown by the 3D island mechanism. During sputtering the defects, appearing between the islands are caused by the stress at the edge of the pores formed at the early stage of film nucleation. These defects and the defective layers, arising at the interfaces MgO – BST and BST – BFO can lead to the breaking of the selection rules and the depolarization of the Raman spectra even though the islands are strictly oriented relative to each other and the substrate. Note, however, that the *XX* and *YY* spectra are similar and contain the same set of lines at the same frequencies, which indicates on the pseudo-tetragonality of the BFO layers.

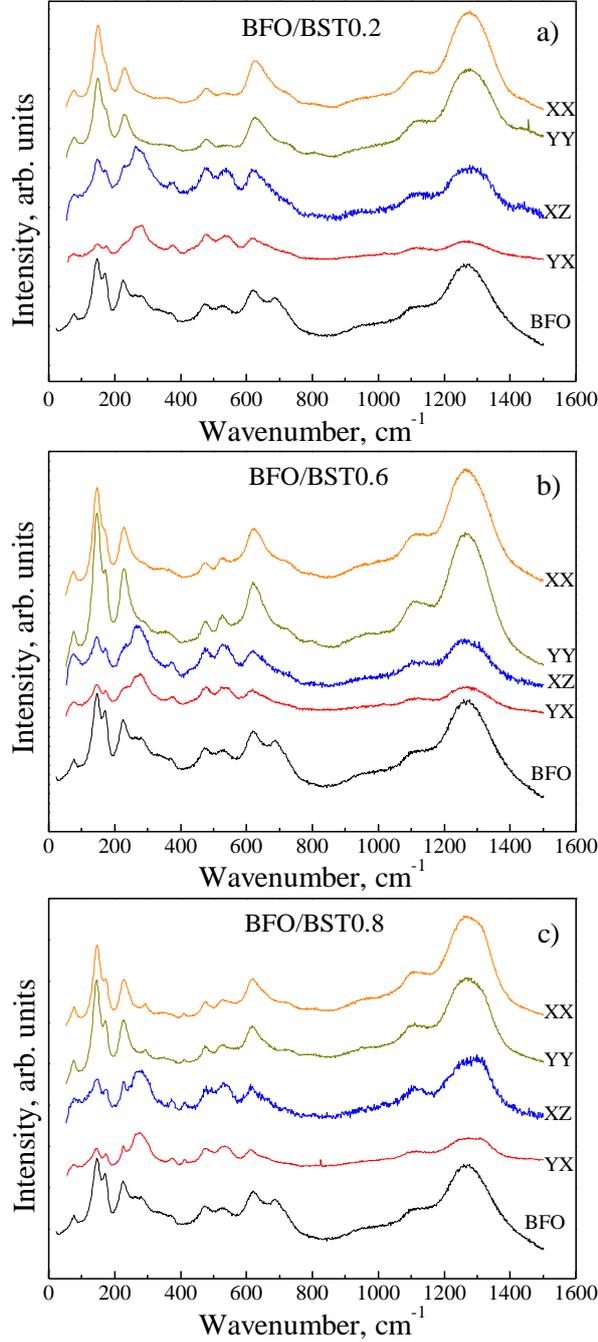

Fig. 8. Polarized Raman spectra of the BFO/BST*x* heterostructures at *x* = 0.2 (*a*), 0.6 (*b*) and 0.8 (*c*), obtained in different scattering geometries at room temperature. For comparison, the Raman spectra of the BFO thin film are shown.

The Raman lines in the spectra of bulk BFO single-crystal, caused by the first-order optical phonons have frequencies less than 600 cm$^{-1}$ [13, 14]. As shown in [32, 33], the band at a frequency of ~620-630 cm$^{-1}$ corresponds to the maximum density of states of the magnon branch at the Brillouin zone boundary. It appears in the Raman spectra because of the breakdown of the selection rules for the wave vector due to the translational symmetry breaking, caused by the defects in the BFO films. The band at 1250-1270 cm$^{-1}$ corresponds to the two-magnon excitations [32]. The intensity of this band significantly higher than the intensity of the two-phonon bands. This can be explained by a resonant process due to the closeness of the energy of the exciting radiation (2.41 eV) and energy of the electronic transition between $e_g$- and $t_{2g}$-orbitals of the $Fe^{3+}$ ions (2-2.5 eV). A necessary condition for such resonance is the exchange interaction between two electrons of the nearest $Fe^{3+}$ ions in the excited state [34, 35]. Using the inelastic neutron scattering technique, the dispersion of the magnon branches over the entire Brillouin zone was measured for the $BiFeO_3$ single crystal [33] and the constant of the exchange interaction (J = 4.38 meV was obtained. This result is consistent with the interpretation of the band around 1250-1270 $cm^{-1}$ as a two-magnon peak.

## 4. Conclusions

The symmetric bi-layered BFO/BST$x$ heterostructures with a total thickness of 200 nm were grown on single-crystal MgO substrates using by rf sputtering. Atomic force microscopy revealed that the BFO layers were grown according to the 3D island growth mechanism, in which the BFO blocks are strictly oriented relative to each other and the crystallographic axes of the MgO substrate. According to x-ray diffraction and Raman spectroscopy data, the BST$x$ layers in the studied heterostructures have a tetragonal symmetry at room temperature. It was shown that the degree of the tetragonal distortion of the BST$x$ layers increases with the increasing of the Ba concentration.

The unit cell parameters of the BFO layers were calculated both in rhombohedral and pseudocubic representations. We found that the BFO layer grown on the paraelectric BST0.2 buffer layer has the largest rhombohedral distortion. The rhombohedral angle of the crystal lattice in the BFO layers decreases when Ba concentration increases in the BST$x$ layers. Decreasing the rhombohedral angle causes the rotation of the polarization vector in the BFO layers. This observation is explained by the effect of the electric field induced by the polarization of BST layers, arising at $x > 0.5$. Such possibility to control the polarization in the BFO thin films using the built-in electric field from the buffer layer open the new way with respect to the previously used method, in which polarization was tuned by changing the two-dimensional stress from substrates. Remarkably, the suggested electrostatic control has a stronger effect on the BFO films than the traditional strain tuning.

**Acknowledgements**

This study was supported by the Russian Foundation for Basic Research (projects No. 16-32-00033 mol_a and No. 17-02-01247a), by H2020-RISE-ENGIMA action (IL and AR). AR thanks grant of President RF No. SP-1359.2016.3.